\documentclass[twoside,twocolumn,9pt]{article}
\usepackage{extsizes}
\usepackage[super,sort&compress,comma]{natbib} 
\usepackage[version=3]{mhchem}
\usepackage[left=1.5cm, right=1.5cm, top=1.785cm, bottom=2.0cm]{geometry}
\usepackage{balance}
\usepackage{mathptmx}
\usepackage{sectsty}
\usepackage{graphicx} 
\usepackage{lastpage}
\usepackage[format=plain,justification=justified,singlelinecheck=false,font={stretch=1.125,small,sf},labelfont=bf,labelsep=space]{caption}
\usepackage{float}
\usepackage{fancyhdr}
\usepackage{fnpos}
\usepackage[english]{babel}
\addto{\captionsenglish}{%
  \renewcommand{\refname}{Notes and references}
}
\usepackage{array}
\usepackage{droidsans}
\usepackage{charter}
\usepackage[T1]{fontenc}
\usepackage[usenames,dvipsnames]{xcolor}
\usepackage{setspace}
\usepackage[compact]{titlesec}
\usepackage[symbol]{footmisc}
\usepackage{hyperref}
\usepackage{hyperref}
\usepackage{amssymb}
\usepackage{siunitx}
\usepackage{upgreek}
\usepackage{threeparttable}

\usepackage{epstopdf}

\graphicspath{{Figures/}}

\definecolor{cream}{RGB}{222,217,201}

\begin{document}

\pagestyle{fancy}
\thispagestyle{plain}
\fancypagestyle{plain}{
\renewcommand{\headrulewidth}{0pt}
}

\makeFNbottom
\makeatletter
\renewcommand\LARGE{\@setfontsize\LARGE{15pt}{17}}
\renewcommand\Large{\@setfontsize\Large{12pt}{14}}
\renewcommand\large{\@setfontsize\large{10pt}{12}}
\renewcommand\footnotesize{\@setfontsize\footnotesize{7pt}{10}}
\makeatother

\renewcommand{\thefootnote}{\fnsymbol{footnote}}
\renewcommand\footnoterule{\vspace*{1pt}%
\color{cream}\hrule width 3.5in height 0.4pt \color{black}\vspace*{5pt}} 
\setcounter{secnumdepth}{5}

\makeatletter 
\renewcommand\@biblabel[1]{#1}            
\renewcommand\@makefntext[1]%
{\noindent\makebox[0pt][r]{\@thefnmark\,}#1}
\makeatother 
\renewcommand{\figurename}{\small{Fig.}~}
\sectionfont{\sffamily\Large}
\subsectionfont{\normalsize}
\subsubsectionfont{\bf}
\setstretch{1.125} 
\setlength{\skip\footins}{0.8cm}
\setlength{\footnotesep}{0.25cm}
\setlength{\jot}{10pt}
\titlespacing*{\section}{0pt}{4pt}{4pt}
\titlespacing*{\subsection}{0pt}{15pt}{1pt}

\fancyfoot{}
\fancyfoot[LO,RE]{\vspace{-7.1pt}\includegraphics[height=9pt]{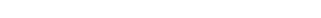}}
\fancyfoot[CO]{\vspace{-7.1pt}\hspace{13.2cm}\includegraphics{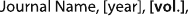}}
\fancyfoot[CE]{\vspace{-7.2pt}\hspace{-14.2cm}\includegraphics{head_foot/RF}}
\fancyfoot[RO]{\footnotesize{\sffamily{1--\pageref{LastPage} ~\textbar  \hspace{2pt}\thepage}}}
\fancyfoot[LE]{\footnotesize{\sffamily{\thepage~\textbar\hspace{3.45cm} 1--\pageref{LastPage}}}}
\fancyhead{}
\renewcommand{\headrulewidth}{0pt} 
\renewcommand{\footrulewidth}{0pt}
\setlength{\arrayrulewidth}{1pt}
\setlength{\columnsep}{6.5mm}
\setlength\bibsep{1pt}

\makeatletter 
\newlength{\figrulesep} 
\setlength{\figrulesep}{0.5\textfloatsep} 

\newcommand{\topfigrule}{\vspace*{-1pt}%
\noindent{\color{cream}\rule[-\figrulesep]{\columnwidth}{1.5pt}} }

\newcommand{\botfigrule}{\vspace*{-2pt}%
\noindent{\color{cream}\rule[\figrulesep]{\columnwidth}{1.5pt}} }

\newcommand{\dblfigrule}{\vspace*{-1pt}%
\noindent{\color{cream}\rule[-\figrulesep]{\textwidth}{1.5pt}} }

\makeatother

\twocolumn[
  \begin{@twocolumnfalse}
{\includegraphics[height=30pt]{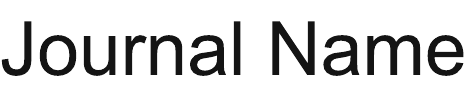}\hfill\raisebox{0pt}[0pt][0pt]{\includegraphics[height=55pt]{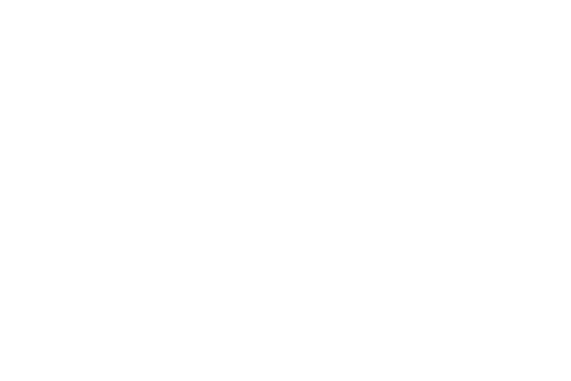}}\\[1ex]
\includegraphics[width=18.5cm]{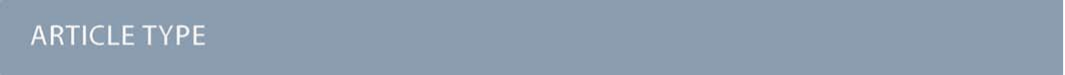}}\par
\vspace{1em}
\sffamily
\begin{tabular}{m{4.5cm} p{13.5cm} }

\includegraphics{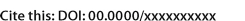} & \noindent\LARGE{\textbf{Judd-Ofelt analysis and coherent spectroscopy of europium-based molecular complexes for quantum information applications}} \\
\vspace{0.3cm} & \vspace{0.3cm} \\

 & \noindent\large{Nicholas Lester Jobbitt,\textit{$^{a}$} Sascha Weis,\textit{$^{a}$} Timo Schweizer,\textit{$^{b}$} Vishnu Unni Chorakkunnath,\textit{$^{a}$} Evgenij Vasilenko,\textit{$^{c}$} Tobias Haas,\textit{$^{b}$} Barbora Brachnakova,\textit{$^{d}$} Senthil Kumar Kuppusamy,\textit{$^{c,d}$} Hartmut Schubert,\textit{$^{b}$} Mario Ruben,\textit{$^{c,d,e}$} Michael Seitz,\textit{$^{b}$} and David Hunger\textit{$^{c, f}$}}  \\

\includegraphics{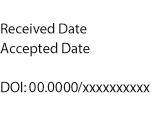} & \noindent\normalsize{Rare-earth-ion based molecular crystals are emerging as a promising platform for quantum information applications, combining long optical coherence times with the chemical tunability of coordination chemistry. Here we present a comprehensive optical characterisation of twelve Eu$^{3+}$-based molecular complexes, with a focus on the coherent $^{5}\text{D}_{0} \longrightarrow \ ^{7}\text{F}_{0}$ transition at $\sim$580 nm. Judd-Ofelt analyses are performed across all complexes, providing insights into ligand-field environments, local point-group symmetries, and the resulting branching ratios of the coherent transition. We find that lower point-group symmetry (C$_1$, C$_2$)  yields higher branching ratios of the $^{5}\text{D}_{0} \longrightarrow \ ^{7}\text{F}_{0}$ transition, exceeding 1\% in several complexes. Optical inhomogeneous linewidths measured at 4.2 K reveal narrow linewidths for some complexes of below ten GHz, indicative of a highly ordered crystalline environment. Optical coherence measurements of one of the complexes, 5-Eu, yields an optical coherence time of $T_{\text{2}}^{\text{opt}} = 2.3~\upmu\mathrm{s}$ at 4.2 K, comparable to the best-performing molecular systems reported to date. Systematic characterisation and quantitative comparison will be key to find structure-property relations guiding rational design of optimised complexes for quantum information applications.} \\

\end{tabular}

 \end{@twocolumnfalse} \vspace{0.6cm}

  ]

\renewcommand*\rmdefault{bch}\normalfont\upshape
\rmfamily
\section*{}
\vspace{-1cm}


\footnotetext{\textit{$^{a}$~Physikalisches Institut, Karlsruhe Institute of Technology, Karlsruhe, Germany.}}
\footnotetext{\textit{$^{b}$~Institute of Inorganic Chemistry, University of Tübingen, Tübingen, Germany.}}
\footnotetext{\textit{$^{c}$~Institute of Quantum Materials and Technologies, Karlsruhe Institute of Technology, Karlsruhe, Germany.}}
\footnotetext{\textit{$^{d}$~Institute of Nanotechnology, Karlsruhe Institute of Technology, Karlsruhe, Germany.}}
\footnotetext{\textit{$^{e}$~Centre Européen de Sciences Quantiques (CESQ), Institut de Science et d’Ingénierie Supramoléculaires (ISIS), Strasbourg, France.}}
\footnotetext{\textit{$^{f}$~Physikalisches Institut, Karlsruhe Institute of Technology, Karlsruhe, Germany. E-mail: david.hunger@kit.edu}}




\section{Introduction}

\begin{figure}[h!]
 \centering
 \includegraphics[width=0.48\textwidth]{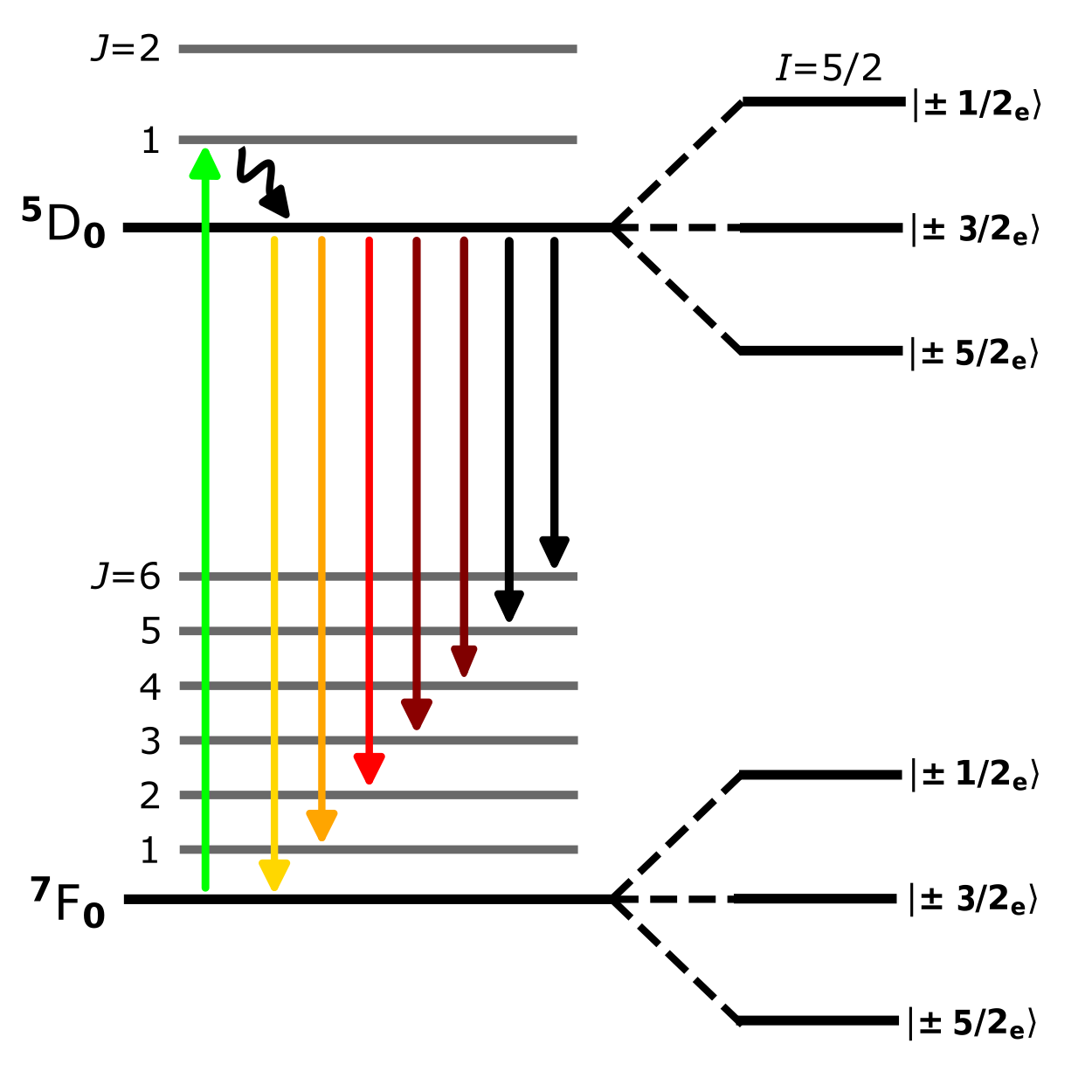}
 \caption{Schematic showing the energy-level system of Eu$^{3+}$. The coherent $^{5}\text{D}_{0} \longrightarrow \ ^{7}\text{F}_{0}$ transition is located at a wavelength of approximately 580 nm. Each state is comprised of three doubly-degenerate hyperfine states due to the $I= 5/2$ nuclear spin of both naturally occurring isotopes of europium.}
 \label{fgr1}
\end{figure}

Rare-earth-ion (REI) based solid-state materials have garnered significant attention in the realisation of quantum computing and quantum networking nodes due to their exceptional optical (4 ms) and spin (18 hours) coherence times \cite{Bottger, Fudong_2025}, connection of optical and microwave spin transitions \cite{Fernandez-Gonzalvo_2015, Probst_2013,Xie_scalable_2025}, and compatibility with current telecommunication infrastructure \cite{Bottger, Fernandez-Gonzalvo_2015,ourari_indistinguishable_2023}. REIs have significant potential for the realisation of high-connectivity quantum processing nodes, and comprehensive studies of REI-based quantum gates and possible protocols can be found in \cite{kinos_designing_2021,kinos_high-connectivity_2022,kinos2021roadmap}. However, traditional REI doped-inorganic crystals are limited by low oscillator strengths and branching ratios of their coherent transitions. Coupled with challenges in scaling such materials, due to their intrinsic random distribution of ions resulting in differing coupling strength between any pairs of ions, it may prove difficult to create scalable quantum registers. REI-based molecular crystals can serve as an attractive alternative to traditional REI doped-inorganic crystals due to their ability to self-assemble into large-scale single-crystalline structures composed of identical copies. Furthermore, fine-tuning the ligand structure of such molecules gives the ability to optimise the oscillator strength and branching ratio, whilst still maintaining good coherence properties. To date, a number of works have been published that focus on REI-based molecular systems for use in quantum information applications, harnessing a broad range of methods ranging from electron paramagnetic resonance (EPR) to transport measurements \cite{Moreno-Pineda_2018, Wernsdorfer_2019}, and recently extending to optical spectroscopy \cite{serrano_ultra_2022, Kuppusamy_2024, Mor_2022, Shin_2024}. Here, we focus on Eu$^{3+}$ as an optically addressable nuclear spin system, due to this ion's highly coherent $^{5}\text{D}_{0} \longrightarrow \ ^{7}\text{F}_{0}$ transition at about 580~nm, depicted in Figure \ref{fgr1}. For both naturally present isotopes of europium, the nuclear spin is $I= 5/2$, resulting in three doubly-degenerate hyperfine states in each level that can serve as qubits. The exceptionally long coherence for this transition derives from the absence of total angular momentum ($J=0$) for both states, which eliminates electronic contributions to the Zeeman effect and makes the transition minimally sensitive to ligand field fluctuations. In addition, the transition has a large energy gap, resulting in the suppression of multi-phonon relaxation, leaving only weak phonon-induced pure dephasing.

The optical properties of multiple Eu$^{3+}$-based molecular complexes have been published \cite{serrano_ultra_2022, Kuppusamy_2023, Schlittenhardt_2024, Kuppusamy_2021, vasilenko2026_BB, Al-Hunaishi_2026}, and the spin coherence of one complex has been investigated \cite{Vasilenko2026-vu}, showing coherence times that are competitive with many inorganic hosts. Although several reviews have been published recently comparing the luminescence properties of Eu$^{3+}$-based molecular complexes \cite{Ebert_2025, Kuppusamy_2024, MOURA_2022, Sedykh_2024}, only two studies have provided an overview of the optical properties of such complexes in the context of quantum information applications \cite{Kuppusamy_2024, Mann2026}.

Here, we provide a comprehensive analysis of twelve Eu$^{3+}$-based molecular complexes shown in Table \ref{tbl1} in the context of quantum information applications, with a focus on the luminescence and coherence properties of the coherent $^{5}\text{D}_{0} \longrightarrow \ ^{7}\text{F}_{0}$ transition. Shown are the short names referred to in this work, their molecular formulae, and the local point-group symmetries about the Eu$^{3+}$ ion. The luminescence properties are complemented with Judd-Ofelt analyses, giving insights into the bonding type, the effective local point-group symmetry about the ion, and the radiative quantum efficiency. 
However, it should be noted that Judd-Ofelt theory does not account for the mixing between states of different $J$ values, and this mixing is crucial to allow the coherent $^{5}\text{D}_{0} \longrightarrow \ ^{7}\text{F}_{0}$ transition which is nominally forbidden by selection rules.

Several factors have been identified that contribute to the strength and branching ratio of the $^{5}\text{D}_{0} \longrightarrow \ ^{7}\text{F}_{0}$ transition \cite{BINNEMANS20151}. First, as a general aspect, a lower point-group symmetry about the Eu$^{3+}$ ion allows for more mixing between different $J$-states and thereby increases the branching ratio. Furthermore, distortions from the determined ideal local point-group symmetries allow for an enhancement of the $^{5}\text{D}_{0} \longrightarrow \ ^{7}\text{F}_{0}$ transition strength, in some cases resulting in the allowance of the transition which should be nominally forbidden. This has been observed in the present study for the case of 7-Eu, 8-Eu, 9-Eu and 11-Eu, where the point-group symmetries were determined by continuous shape measures to be distorted D$_{\text{4d}}$, D$_{\text{4d}}$, D$_{\text{4d}}$ and  D$_{\text{5h}}$, resulting in effective symmetries of  C$_{\text{1}}$, C$_{\text{1}}$, C$_{\text{2v}}$ and C$_{\text{2h}}$ respectively \cite{serrano_ultra_2022, Kuppusamy_2021}.

Second, the magnitude of the ligand-field interaction is determined not only by the formal site symmetry but also by the charge, polarisability, covalency, distance, and spatial arrangement of the coordinating atoms. These factors determine the crystal-field parameters and therefore the extent of mixing between different $J$-manifolds. Oxygen-rich coordination environments can produce substantial ligand-field interactions around Eu$^{3+}$, although the resulting splitting and $J$-mixing cannot be inferred from donor-atom identity alone \cite{BINNEMANS20151}. A large ligand polarisability, such as those provided by cryptates (1-Eu and 2-Eu), was observed to lead to an enhancement of the oscillator strength \cite{Haas_2026}. Furthermore, different coordinating species can provide an axial ligand field even at higher point symmetry and thereby induce the transition, as observed here, e.g., for 12-Eu (C$_{\text{3v}}$) in contrast to the absence of the transition in 3-Eu (C$_{\text{3}}$).

Third, binuclear complexes (8-Eu and 11-Eu) have also been observed to yield a higher branching ratio relative to their mononuclear counterparts \cite{Kuppusamy_2021, vasilenko2026_BB}. This may be attributed to additional polarisability arising from the interactions between identical ions. 

Finally, also a low-lying ligand-to-Eu$^{3+}$ charge transfer state that admixes to the $4f$ levels can enhance the $^{5}\text{D}_{0} \longrightarrow \ ^{7}\text{F}_{0}$ transition.

\begin{table}[h!]
\small
  \caption{\ Molecular complexes investigated in this study, including their point-group symmetries and previous works investigating these molecular complexes.}
  \label{tbl1}
  \begin{tabular*}{0.48\textwidth}{@{\extracolsep{\fill}}llll}
    \hline
    Name & Complex & Symmetry & Refs.\\
    \hline
    1-Eu & \ce{[Eu(bipyO2)(phen)2](TFA)3} & C$_{\text{2}}$ & \cite{Haas_2026}\\
    2-Eu & \ce{[Eu(bipyO2)2(phen)](TFA)3} & C$_{\text{2}}$ & \cite{Haas_2026}\\
    3-Eu & \ce{[Eu(tpip)3]} & C$_{\text{3}}$ & \cite{PIETRASZKIEWICZ_2012}\\
    4-Eu & \ce{[Eu(tpip)3(phen)]} & C$_{\text{1}}$ & \cite{PIETRASZKIEWICZ_2012}\\
    5-Eu & \ce{[Eu(dbm)3(bipy)]} & C$_{\text{1}}$& \cite{Wong_2017}\\
    6-Eu & \ce{[Eu(dbm)3(bipyO2)]} & C$_{\text{1}}$ & This study\\
    7-Eu & \ce{[Eu(btfa)3(bipy)]} &  D$_{\text{4d}}$ (C$_{\text{1}}$) & \cite{Batista_1998, vasilenko2026_BB}\\
    8-Eu & \ce{[Eu2(btfa)6(bpim)]} & D$_{\text{4d}}$ (C$_{\text{1}}$) & \cite{Ilmi_2020, vasilenko2026_BB}\\
    9-Eu & \ce{[Eu(BA)4](pip)} & D$_{\text{4d}}$ (C$_{\text{2v}}$) & \cite{serrano_ultra_2022, Vasilenko2026-vu}\\
    10-Eu & \ce{[Eu(dpphen)(NO3)3]} & C$_{\text{2v}}$* & \cite{Schlittenhardt_2024}\\
    11-Eu & \ce{[Eu2Cl6(4-picNO)6].2H2O} & D$_{\text{5h}}$ (C$_{\text{2h}}$) & \cite{Kuppusamy_2021}\\
    12-Eu & \ce{[Eu(trensal)]}  & C$_{\text{3v}}$ & \cite{Kuppusamy_2023} \\
    \hline
  \end{tabular*}
  bipyO2 = 2,2'-bipyridine-N,N'-dioxide, phen = 1,10-phenanthroline, TFA = trifluoroacetate anion, tpip = N-(Diphenylphosphinyl)-P,P-diphenylphosphinic amide, dbm = 1,3-Diphenylpropane-1,3-dione, bipy = 2,2'-bipyridyl, btfa = 4,4,4-Trifluoro-1-phenyl-1,3-butanedione, bpim = 2,2'-bipyrimidyl, BA = 1-phenylbutane-1,3-dione, pip = piperidinium, dpphen = 2,9-bis(pyrazol-1-yl)-1,10-phenanthroline, picNO = 4-picoline-N-oxide, trensal = deprotonised 2,2,2-tris(salicylideneimino)trimethylamine.
\end{table}

\footnotetext{* For 10-Eu, the symmetry observed through single-crystal X-ray diffraction is distorted C$_{\text{2v}}$ for both polymorphs; however, the degree of distortion, and therefore the resulting effective symmetries (either C$_{\text{n}}$, C$_{\text{nv}}$, or C$_{\text{s}}$) are uncertain.}

Many pairs of the molecular complexes investigated in this work are structurally and chemically related, namely 1-Eu and 2-Eu, 3-Eu and 4-Eu, 5-Eu and 6-Eu, and 7-Eu and 8-Eu. For this reason we pair these complexes together in all relevant figures. For consistency and formatting, we also pair 9-Eu and 10-Eu in addition to 11-Eu and 12-Eu despite no chemical or structural relations between these complexes.

Most of the complexes presented here are mononuclear, however, 8-Eu and 11-Eu are binuclear. While 8-Eu has been shown to be optically degenerate as only a single $^{5}\text{D}_{0} \longrightarrow \ ^{7}\text{F}_{0}$ optical transition has been observed \cite{vasilenko2026_BB}, 11-Eu exhibited multiple sub-sites \cite{Kuppusamy_2021}, all with various intensities. Therefore, all values presented in this study for 11-Eu represent an average over all sub-sites.

\section{Experimental}

\subsection{Sample preparation}

Mononuclear and binuclear complexes are generally synthetised by coordinating Eu$^{3+}$ salts with organic chelating and bridging ligands in the appropriate solvents. Ligands containing acidic protons, such as beta-diketones (\ce{HHBA}, \ce{Hbtfa}, \ce{Hbtfa}), imidodiphosphinates (\ce{Htpip}) or tripodal Schiff bases (\ce{H3trensal}) are added to in situ stoichiometric amounts of a base (e.g. triethylamine or \ce{NaOH}) to achieve full deprotonation and facilitate coordination to the Eu$^{3+}$ center. Neutral N-heterocyclic ligands (e.g. bipyridine, phenanthroline, pyridine-N-oxides, dpphen) are added either simultaneously or sequentially to saturate the Eu$^{3+}$ coordination sphere (CN = 6-8) and replace coordinated water molecules. The resulting complexes are isolated as microcrystalline powders via direct precipitation, slow solvent evaporation, or recrystallisation from various solvent mixtures. The europium cryptates are obtained by metal exchange of the corresponding Na-cryptates, followed by HPLC purification. All measurements were performed on microcrystalline powders in the solid-state. All complexes have been fully characterised in previous works, except of the newly synthesised 6-Eu. The synthetic procedure and crystallographic characterisation of 6-Eu are discussed in the supporting information, S1.

\subsection{Experimental methods}

The fluorescence spectra were measured at room temperature using the Princeton instruments Acton SP2500i (150 grooves/mm, 500 nm blaze) coupled to a PIXIS 2k eXcelon charge-coupled device (CCD). The setup was both wavelength and intensity calibrated using the Intellical Hg and Visible - NIR calibration lamp, respectively. Excitation was provided using a Laser Quantum Ventus CW laser at 532 nm.

Optical lifetimes were measured at room temperature using a Thorlabs Si avalanche photo-diode APD130A2/M. Excitation was provided using a Laser Quantum Ventus CW laser at 532 nm, which was pulsed with a Gooch \& Housego  acoutic-optical modulator AOMO 3200-125 to provide an excitation pulse of length 20$~\upmu\mathrm{s}$.

The inhomogeneous linewidth, spectral hole burning, free-induction decay and photon-echo measurements were performed in a ferrule-based setup with multimode excitation and collection fibres. The ferrule was placed into liquid helium at a temperature of 4.2 K. A Sirah Matisse 2 DX dye laser with a linewidth < 100 kHz provided the excitation. All pulse sequences were executed using a Gooch \& Housego acoutic-optical modulator AOMO 3200-1215.

\section{Results and Discussion}

\subsection{Luminescence}

\begin{figure*}[h!]
 \centering
 \includegraphics[width=1\textwidth]{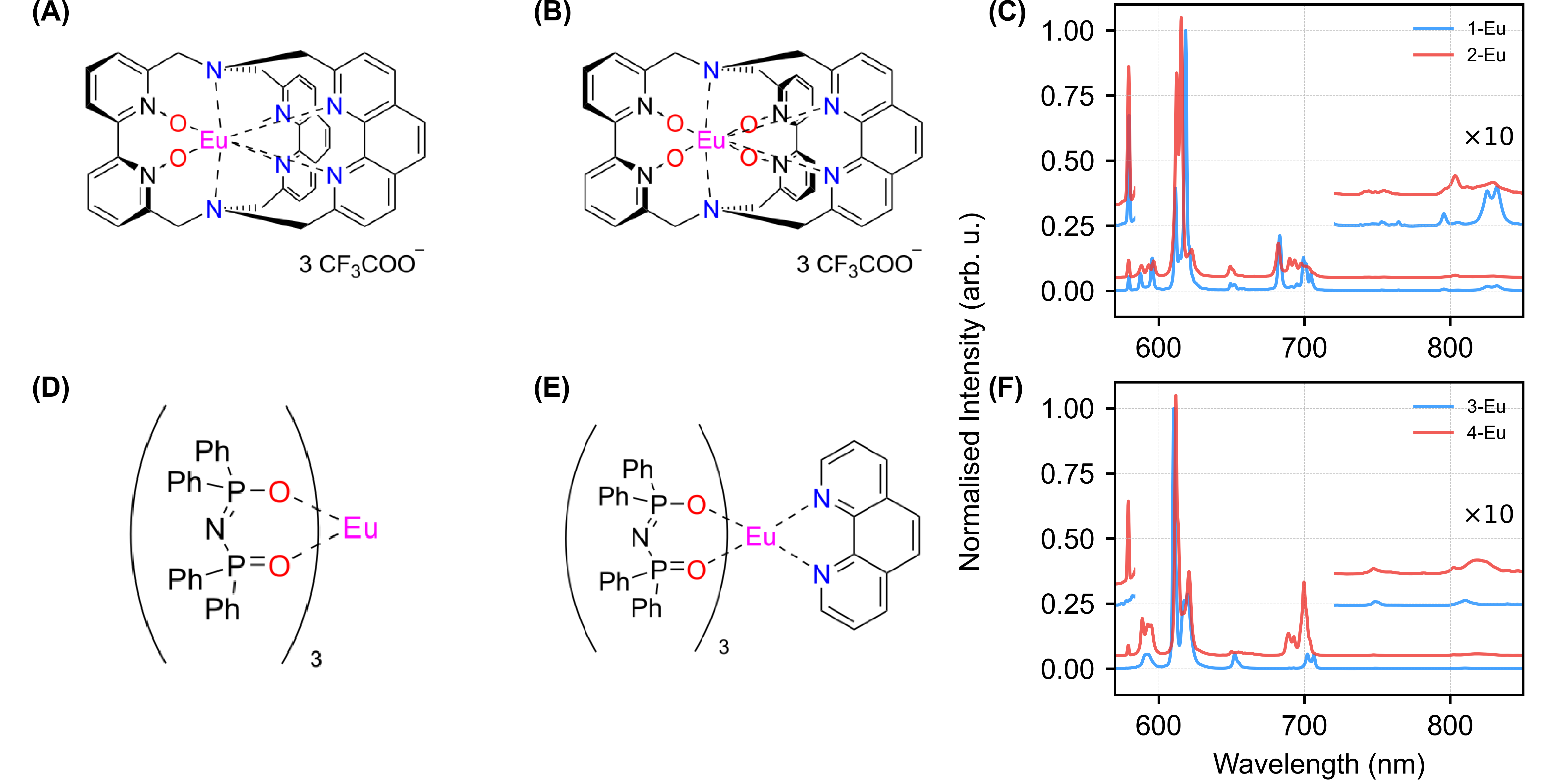}
 \caption{Molecular structures of: (A) 1-Eu, (B) 2-Eu, (D) 3-Eu and (E) 4-Eu, and the measured room temperature fluorescence spectra of (C) 1-Eu and 2-Eu, and (F) 3-Eu and 4-Eu. Excitation was provided at 532 nm. Insets with $\times10$ zoom are provided showing the coherent $^{5}\text{D}_{0} \longrightarrow \ ^{7}\text{F}_{\text{0}}$ transition located at $\sim$580 nm in addition to the $^{5}\text{D}_{0} \longrightarrow \ ^{7}\text{F}_{\text{5, 6}}$ transitions located between 720 and 850 nm.}
 \label{fgr2}
\end{figure*}

\begin{figure*}[h!]
 \centering
 \includegraphics[width=1\textwidth]{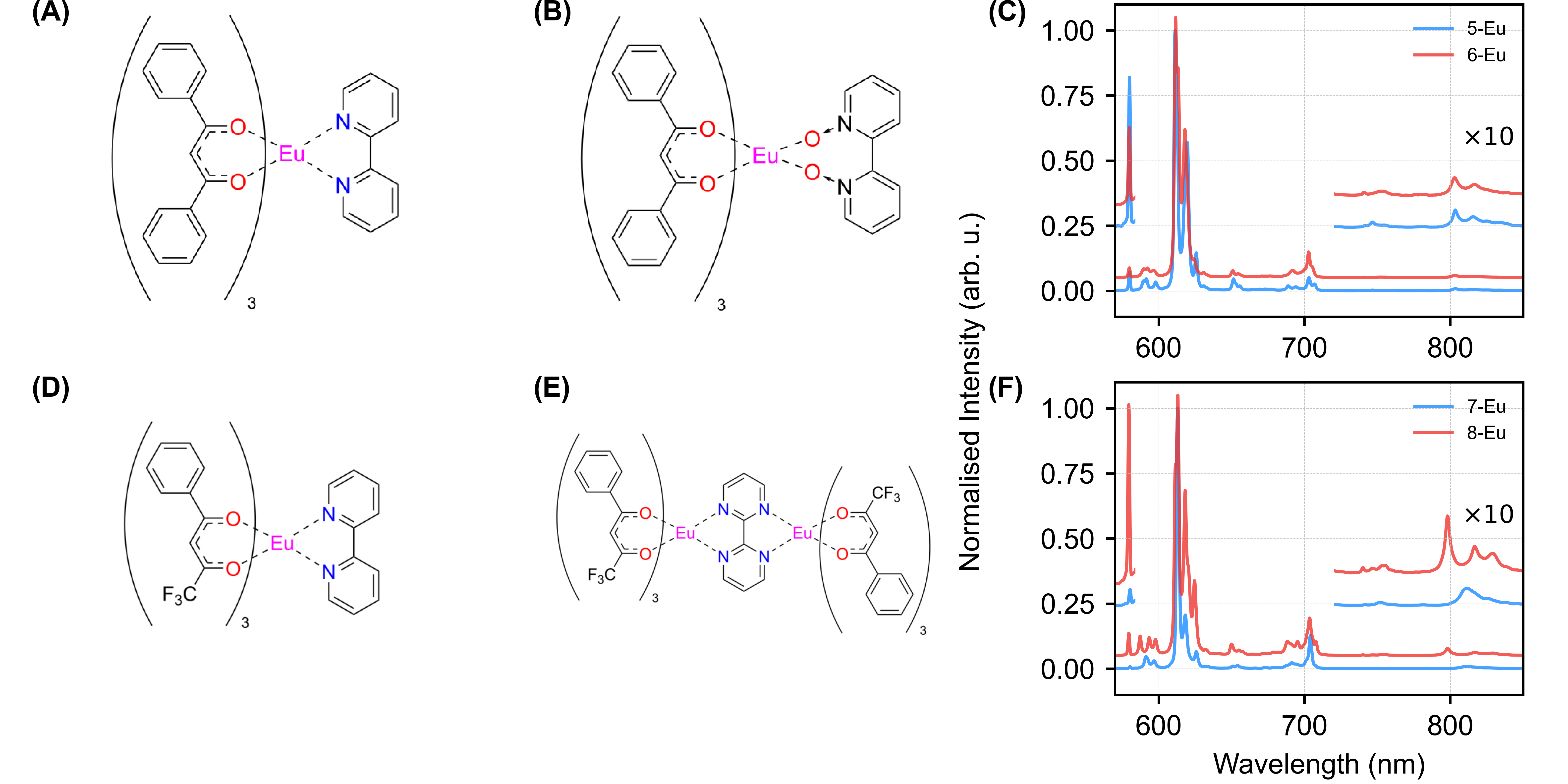}
 \caption{Molecular structures of: (A) 5-Eu, (B) 6-Eu, (D) 7-Eu and (E) 8-Eu, and the measured room temperature fluorescence spectra of (C) 5-Eu and 6-Eu, and (F) 7-Eu and 8-Eu. Excitation was provided at 532 nm. Insets with $\times10$ zoom are provided showing the coherent $^{5}\text{D}_{0} \longrightarrow \ ^{7}\text{F}_{\text{0}}$ transition located at $\sim$580 nm in addition to the $^{5}\text{D}_{0} \longrightarrow \ ^{7}\text{F}_{\text{5, 6}}$ transitions located between 720 and 850 nm.}
 \label{fgr3}
\end{figure*}

\begin{figure*}[h!]
 \centering
 \includegraphics[width=1\textwidth]{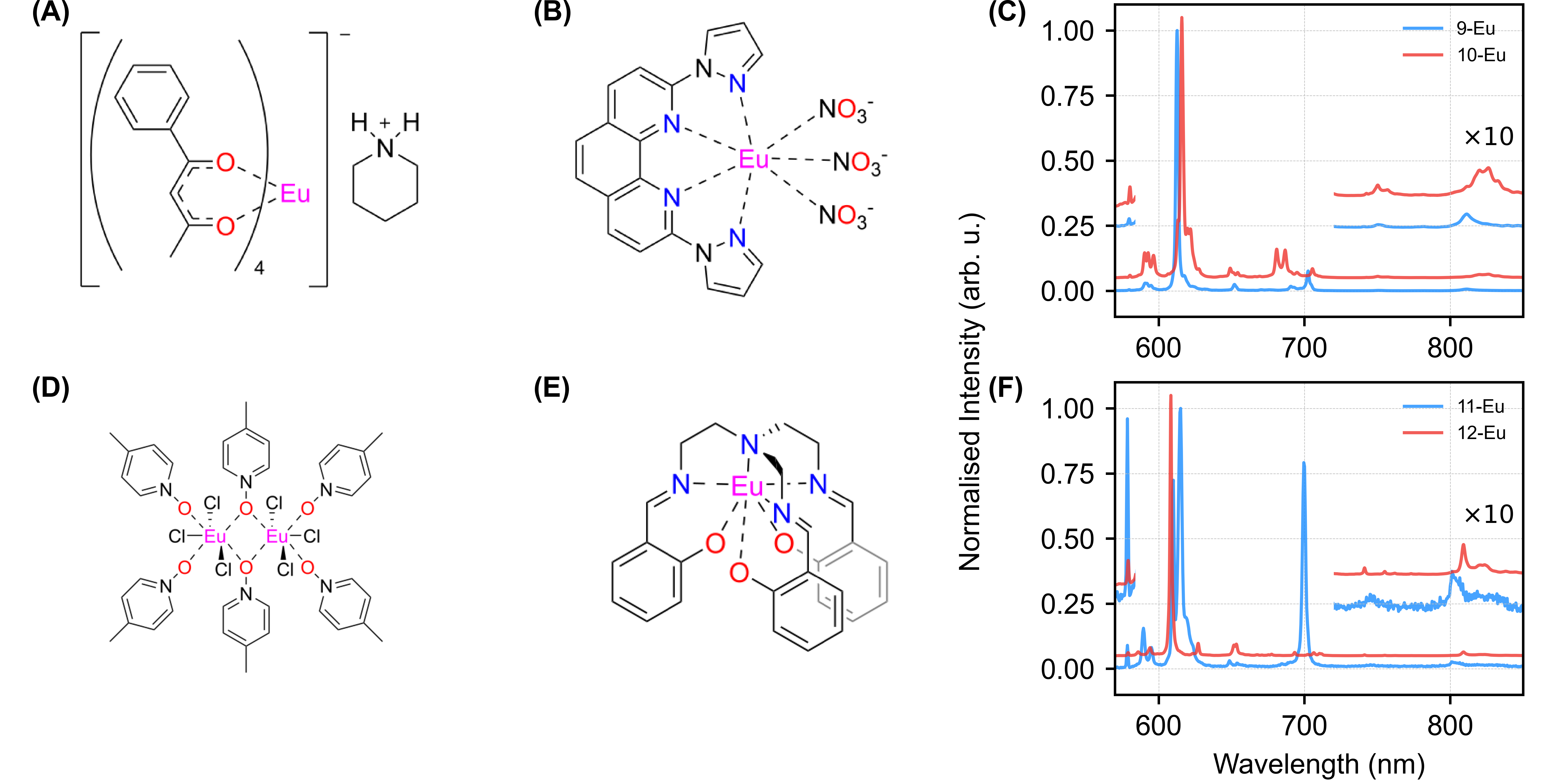}
 \caption{Molecular structures of: (A) 9-Eu, (B) 10-Eu, (D) 11-Eu and (E) 12-Eu, and the measured room temperature fluorescence spectra of (C) 9-Eu and 10-Eu, and (F) 11-Eu and 12-Eu. Excitation was provided at 532 nm. Insets with $\times10$ zoom are provided showing the coherent $^{5}\text{D}_{0} \longrightarrow \ ^{7}\text{F}_{\text{0}}$ transition located at $\sim$580 nm in addition to the $^{5}\text{D}_{0} \longrightarrow \ ^{7}\text{F}_{\text{5, 6}}$ transitions located between 720 and 850 nm.}
 \label{fgr4}
\end{figure*}

The molecular structures and obtained room temperature fluorescence spectra for all molecules listed in Table \ref{tbl1} are shown in Figures \ref{fgr2} -\ref{fgr4}. The complex 11-Eu was not available for the present experiments such that it could not be measured and was instead reproduced, with permission, from the authors of ref. \cite{Kuppusamy_2021}. An excitation wavelength of 532 nm was used to 
excite the $^{7}\text{F}_{1} \longrightarrow \ ^{5}\text{D}_{1}$ transition. 
The excitation then non-radiatively decays to the $^{5}\text{D}_{0}$ level before fluorescing to give the well-documented $^{5}\text{D}_{0} \longrightarrow \ ^{7}\text{F}_{J}$ ($J = 0, \cdots, 6$) transitions visible in Figures \ref{fgr2} - \ref{fgr4}. The dominant feature throughout all spectra is the hypersensitive forced electric-dipole $^{5}\text{D}_{0} \longrightarrow \ ^{7}\text{F}_{\text{2}}$ transition located at $\sim$615 nm. The ratio between this transition and the ligand-field insensitive magnetic-dipole $^{5}\text{D}_{0} \longrightarrow \ ^{7}\text{F}_{\text{1}}$ transition located at $\sim$590 nm gives information about the local symmetry of the REI as the $^{5}\text{D}_{0} \longrightarrow \ ^{7}\text{F}_{\text{2}}$ transition is forbidden under inversion symmetry. The predominance of this transition indicates that all molecular complexes have a low local point-group symmetry of the REI. Of particular interest for this work is the coherent $^{5}\text{D}_{0} \longrightarrow \ ^{7}\text{F}_{\text{0}}$ transition located at $\sim$580 nm. While this transition is particularly weak for 7-Eu, 9-Eu and 10-Eu, it is much stronger in the case of 1-Eu, 2-Eu, 5-Eu, 8-Eu, and 11-Eu. In the case of 3-Eu, this transition is completely forbidden due to the higher point-group symmetry (C$_{3}$) of the complex. Each $^{7}\text{F}_{J}$ multiplet has a degeneracy of $2J + 1$ which is partially or fully lifted by the local point-group symmetry of the Eu$^{3+}$ ion, with symmetries of C$_{\text{2v}}$ and below fully lifting this degeneracy \cite{BINNEMANS20151}. Inspecting the spectra, there is clear evidence that every complex is fully non-degenerate (all $2J+1$ spectral lines visible), with the exception of 3-Eu, 7-Eu, and 11-Eu, which by studying the $^{7}\text{F}_{\text{1}}$\big($^{7}\text{F}_{\text{2}}$\big) multiplets, a total of 2(3), 2(4), 2(3) spectral lines were found for 3-Eu, 7-Eu, and 11-Eu respectively. This is consistent with their respective trigonal, tetragonal and pentagonal symmetries, determined through continuous shape measures.

\begin{figure*}[h!]
 \centering
 \includegraphics[width=1\textwidth]{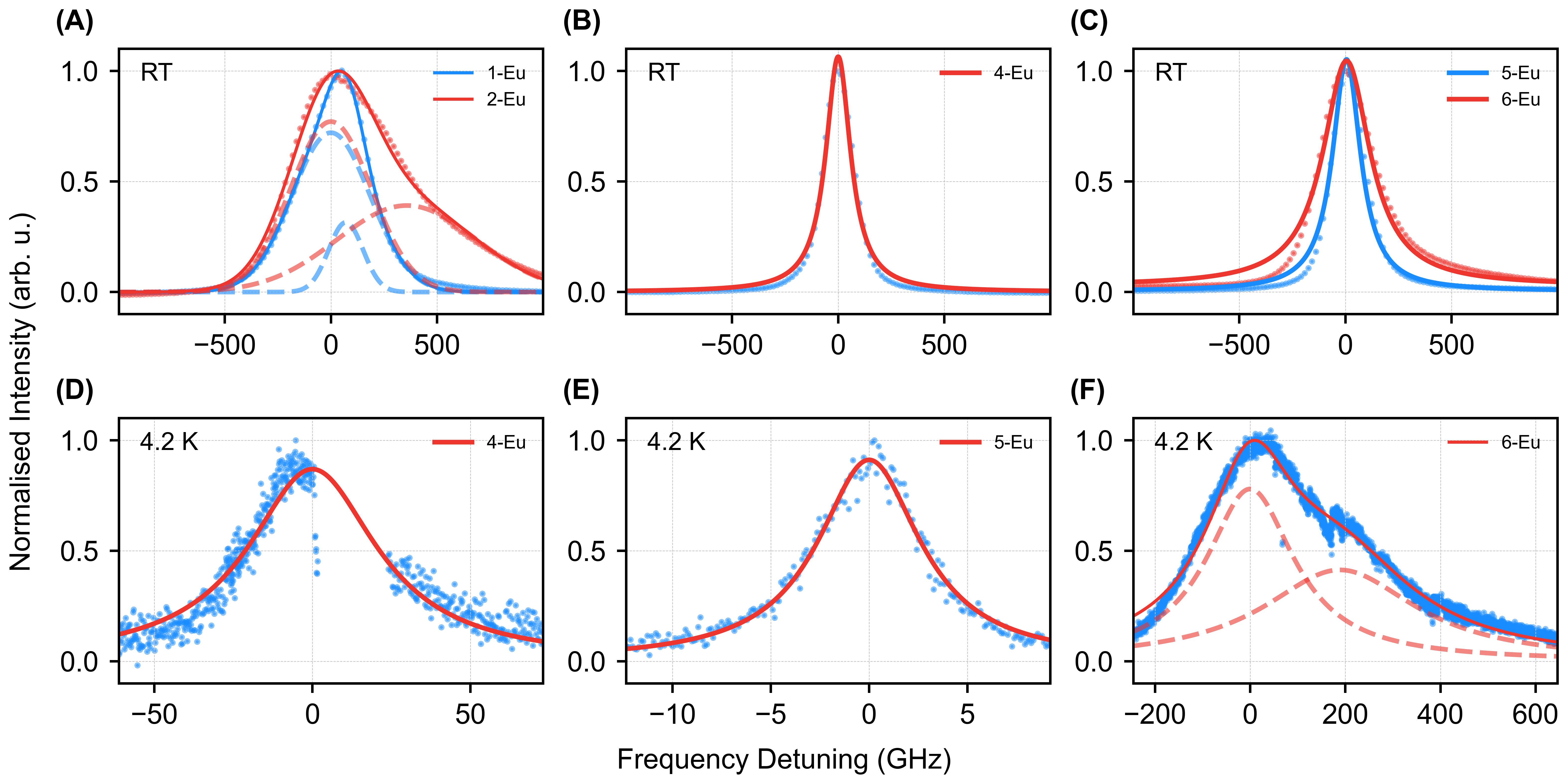}
 \caption{(A) - (C) Inhomogeneous linescans at room temperature of (A) 1-Eu and 2-Eu, (B) 4-Eu, and (C) 5-Eu and 6-Eu. (D) - (F) Inhomogeneous linescans at 4.2 K of (D) 4-Eu, (E) 5-Eu and (F) 6-Eu are also presented. Lorentzian fits were performed on all measurements to extract the optical inhomogeneous linewidths. Double Lorentzian fits were performed on the room temperature measurements of 1-Eu and 2-Eu and the 4.2 K measurements of 6-Eu, whereas all remaining measurements were fitted with a single Lorentzian. The resulting fitted values are presented in Table \ref{tbl2}.}
 \label{fgr5}
\end{figure*}

In addition to the information gained on the local point-group symmetry from the overall spectrum, it is also possible to gauge the degree of crystalline homogeneity (strain and impurities) of the investigated material through measuring the inhomogeneous linewidth of the $^{5}\text{D}_{0} \longrightarrow \ ^{7}\text{F}_{\text{0}}$ transition. The inhomogeneous linewidth of some molecular complexes discussed here have been reported at cryogenic temperatures \cite{serrano_ultra_2022, Kuppusamy_2021, Schlittenhardt_2024, Kuppusamy_2023}. Figure \ref{fgr5} depicts the inhomogeneous linewidth of the molecular complexes yet to be reported, 1-Eu through 6-Eu, with the exception of 3-Eu due to no observable $^{5}\text{D}_{0} \longrightarrow \ ^{7}\text{F}_{\text{0}}$ transition. Panels (A), (B) and (C) show room temperature measurements for 1-Eu and 2-Eu, 4-Eu, and 5-Eu and 6-Eu respectively. 1-Eu and 2-Eu were both found to have two broad inhomogeneous lines of widths 432(2) and 191(4) GHz for 1-Eu and 449(14) and 774(44) GHz for 2-Eu. The existence of two inhomogeneous lines is indicative of the presence of two separate environments in which the Eu$^{3+}$ ion can be situated. This has been seen before in, for example, 10-Eu, due to the presence of two distinct polymorphs \cite{Schlittenhardt_2024} and the presence of multiple sub-sites in the case of 11-Eu \cite{Kuppusamy_2021}. Additionally, the broad inhomogeneous linewidths ($\gg 100$ GHz) measured for both of these complexes is indicative of a disordered or elevated strain environment. While a large inhomogeneous line does not necessarily result in low coherence, it has been noted that this trend is generally true in the case of REI-based molecular complexes. This indicates that static variations in the local crystal-field environment are accompanied by dynamic fluctuations, which contribute to optical dephasing. For this reason, 1-Eu and 2-Eu were not investigated at cryogenic temperatures. 4-Eu, 5-Eu and 6-Eu were all well fitted with single Lorentzians of widths 142(2), 161(2) and 271(4) GHz respectively. A slight asymmetry was noted for 6-Eu, which became visible as two distinct peaks when measured at 4.2 K (see below). All measured room temperature values are well above the spectral limit of the spectrometer used of 36 GHz at 580 nm, determined through the measurement of a narrow-band laser source. Figure \ref{fgr5} (D), (E), and (F) show the linewidths of 4-Eu, 5-Eu, and 6-Eu respectively, measured at a temperature of 4.2 K. 5-Eu shows a very narrow inhomogeneous linewidth of 6.4(0.2) GHz at 4.2 K, very similar to what was measured for 9-Eu \cite{serrano_ultra_2022}, and narrower than 7-Eu, 8-Eu \cite{vasilenko2026_BB}, 10-Eu \cite{Schlittenhardt_2024}, and 12-Eu \cite{Kuppusamy_2023}. Such a narrow linewidth represents a very well-ordered and low strain crystalline environment. 6-Eu shows two broad optical inhomogeneous lines of widths 225(1) and 390(10) GHz, separated by $\sim$200 GHz. Like what was observed for 1-Eu and 2-Eu, the presence of two inhomogeneous lines is indicative of the presence of two separate environments in which the Eu$^{3+}$ ion can be situated. 4-Eu also shows a narrow optical inhomogeneous linewidth of 50(2) GHz at 4.2 K. The fitted transition wavelengths ($\lambda$) and inhomogeneous linewidths ($\Gamma_{\text{inh}}^{\text{opt}}$) of the coherent $^{5}\text{D}_{0} \longrightarrow \ ^{7}\text{F}_{\text{0}}$ transition measured at room temperature and 4.2 K are summarised in Table \ref{tbl2}.

\begin{table}[h!]
\small
  \caption{\ Transition wavelengths ($\lambda$) and optical inhomogeneous linewidths ($\Gamma_{\text{inh}}^{\text{opt}}$) measured at room temperature and at 4.2 K of the coherent $^{5}\text{D}_{0} \longrightarrow \ ^{7}\text{F}_{\text{0}}$ transition for all previously unpublished complexes. Values marked with a '---' were not be measured.}
  \label{tbl2}
  \begin{tabular*}{0.48\textwidth}{@{\extracolsep{\fill}}lccc}
    \hline
    Complex & $\lambda$ [nm] & $\Gamma_{\text{inh}}^{\text{opt}}$(RT) [GHz] & $\Gamma_{\text{inh}}^{\text{opt}}$(4.2 K) [GHz] \\
    \hline
    1-Eu & 579.312 & $434(2)$ \& $191(2)$ & --- \\
    2-Eu & 579.294 & $449(14)$ \& $774(44)$ & --- \\
    3-Eu & Not visible & --- & --- \\
    4-Eu & 579.982 & $142(2)$ & $50(2)$ \\
    5-Eu & 580.408 & $161(2)$ & $6.4(0.2)$\\
    6-Eu & 580.734 & $271(4)$ & $225(1)$ \& $390(10)$\\
    \hline
  \end{tabular*}
\end{table}

\begin{figure*}[h!]
 \centering
 \includegraphics[width=1\textwidth]{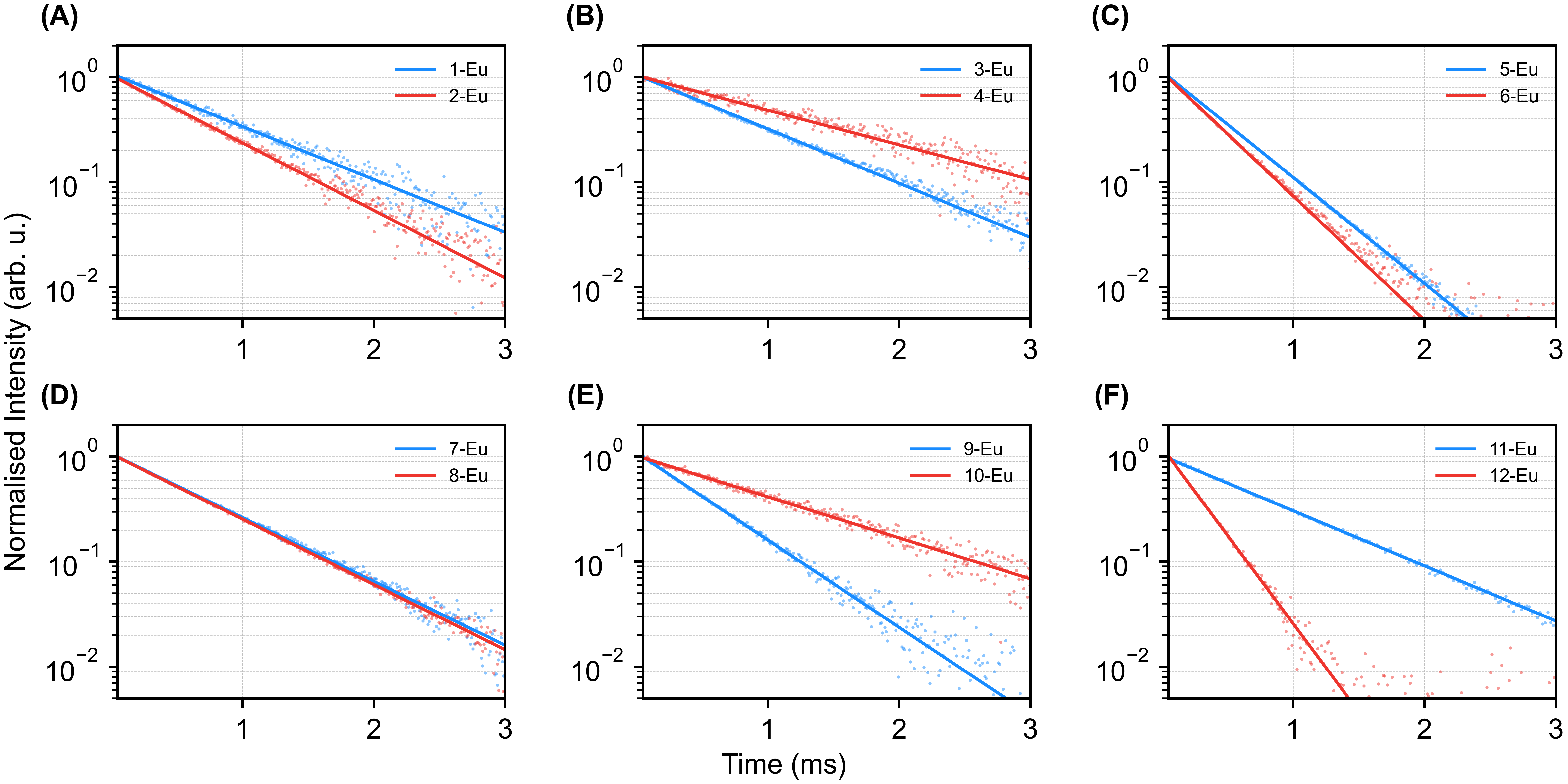}
 \caption{Observed room temperature optical lifetimes $T_{\text{1}}^{\text{opt}}$ with an excitation wavelength of 532 nm for: (A) 1-Eu and 2-Eu, (B) 3-Eu and 4-Eu, (C) 5-Eu and 6-Eu, (D) 7-Eu and 8-Eu, (E) 9-Eu and 10-Eu, and (F) 11-Eu and 12-Eu. Single exponentials were used to fit the data. The fitted values and related uncertainties are presented in Table \ref{tbl3}.}
 \label{fgr6}
\end{figure*}

Optical lifetimes were measured at room temperature with 532 nm excitation for all molecular complexes and the resulting fitted decay curves are shown in Figure \ref{fgr6}. All lifetimes were fitted with a single-exponential. The optical lifetime measured for 9-Eu matches very closely to that previously reported (0.54 ms) \cite{serrano_ultra_2022}, whereas for 10-Eu and 12-Eu there are significant deviations from the previously reported values (1.425 ms and 0.354 ms respectively) \cite{Schlittenhardt_2024, Kuppusamy_2023}. Previously reported values for the optical lifetimes of 7-Eu (0.842 ms) and 8-Eu (0.785 ms and 0.762 ms) also deviate slightly from the values reported here \cite{Ilmi_2020, vasilenko2026_BB}. The observed lifetimes reported here are consistently lower than those previously reported. This is expected as the measurements reported here were taken at room temperature while previously reported values were measured at cryogenic temperatures. Higher temperatures allow for the possibility of excitation transfer to the ligand, excitation hopping to quenching sites, and multi-phonon processes, resulting in a lower lifetime. The largest proportional reduction was measured for 12-Eu (over 25\%) \cite{Kuppusamy_2023}. This is consistent with the comparably low quantum efficiency as calculated below (see Table \ref{tbl3}), and indicates a higher degree of temperature dependent non-radiative pathways. The finite quantum efficiencies of europium-based molecular complexes have been attributed to high energy CH vibrational modes \cite{Ebert_2025}.

\subsection{Judd-Ofelt analyses}
Judd-Ofelt analyses links spectral intensities of emission lines to the chemical and structural surrounding of the Eu$^{3+}$ ion, thereby giving insights in the bonding type, the local point-group symmetry about the ion, and the local ligand-field environment. We note, however, that such an approach does not account for the $J$-mixing present in such systems, pertinent for the $^{5}\text{D}_{0} \longrightarrow \ ^{7}\text{F}_{0}$ transition of interest as this $J = 0 \rightarrow 0$ transition is forbidden by selection rules and therefore Judd-Ofelt theory. To fully account for this, more comprehensive analyses such as Ab initio density-functional theory (DFT) or a crystal-field approach is required \cite{Iyer_2026, Chilton_2025, Morrillo_2024, Jobbitt_2021, jobbitt2022zeeman}. Yet, these approaches are computationally intensive and require a large amount of input data, especially for low point-group symmetry environments, such as those present here. Following the methods of Binnemans \cite{BINNEMANS20151}, the Judd-Ofelt parameters are defined as:

\begin{equation}
  \Omega_{J} \ (J = 2, 4, 6) = \frac{D_{\text{MD}}\tilde{\nu}_{1}^{3}}{e^{2}\tilde{\nu}_{J}^{3}U^{(J)}} \cdot \frac{9n_{1}^{3}}{n_{J}(n_{J}^{2}+2)^{2}}\cdot \frac{A_{J}}{A_{1}}.
  \label{Eq1}
\end{equation}

$D_{\text{MD}}$ is the dipole-strength of the magnetic-dipole  $^{5}\text{D}_{0} \longrightarrow \ ^{7}\text{F}_{1}$ transition, which is $9.6\cdot 10^{-42}$ esu$^2$ cm$^2 \;$ \cite{Martinus}. $\tilde{\nu}_{1}$ and $\tilde{\nu}_{J}$ are the wavenumbers of the barycentres of the $^{5}\text{D}_{0} \longrightarrow \ ^{7}\text{F}_{1}$ and $^{5}\text{D}_{1} \longrightarrow \ ^{7}\text{F}_{J}$ transitions respectively, with corresponding wavelength-dependant refractive indices $n_{1}$ and $n_{J}$. A recent study determined the refractive indices for a number of Eu$^{3+}$-based molecular complexes to be between 1.64 and 1.71 at a wavelength of 589 nm \cite{Sedykh_2024}. Therefore, in this study we used the average of these values ($n = 1.68$) for the refractive index at all wavelengths, and their standard deviation as its uncertainty ($\sigma_n = 0.03$). It is noted that previous studies often use a value of $n = 1.5$ \cite{Ebert_2025}. $U^{(J)}$ are the squared reduced matrix elements and are set to $U^{(2)} = 0.0032$, $U^{(4)} = 0.0023$ and $U^{(6)} = 0.0002$ \cite{Carnall}. Finally, $A_{1}$ and $A_{J}$ are the integrated intensities of the $^{5}\text{D}_{0} \longrightarrow \ ^{7}\text{F}_{1}$ and $^{5}\text{D}_{0} \longrightarrow \ ^{7}\text{F}_{J}$ transitions respectively. With corresponding wavelength ranges around $^5D_0 \rightarrow ^7F_1: (581 - 602)$ nm, $^5D_0 \rightarrow ^7F_2: (605 - 630)$ nm, $^5D_0 \rightarrow ^7F_4: (665 - 720)$ nm and $^5D_0 \rightarrow ^7F_6: (790 - 863)$ nm.

The radiative transition rates are computed following \cite{dacanin}:

\begin{equation}
    I(\Psi J, \Psi' J') = \frac{64\pi^4 \tilde{\nu}_J^3}{3h(2J_{\text{initial}}+1)}\left[\frac{n_{J}(n_{J}^2+2)^2}{9}D_{ED} + n_{1}^3D_{MD}\right],
    \label{EqRadRates}
\end{equation}

where $D_{ED}$ is the strength of the electric dipole transitions, with elementary charge $e$:

\begin{equation}
    D_{ED} = e^2 \sum_{J=2,4,6} \Omega_J U^{(J)},
    \label{Eq3}
\end{equation}

resulting in radiative lifetimes:

\begin{equation}
    T_{\text{1}}^{\text{rad}} = \frac{1}{\sum I_{J\in [1,2,4,6]}}.
    \label{EqRadLife}
\end{equation}

The Judd-Ofelt parameters were calculated using Equation \ref{Eq1} from the obtained spectra shown in Figures \ref{fgr2} - \ref{fgr4} and are presented in Table \ref{tbl3}, along with the calculated radiative lifetimes (Equation \ref{EqRadLife}), the observed optical lifetimes presented in Figure \ref{fgr5} and the quantum efficiencies, $\eta = T_{\text{1}}^{\text{opt}} / T_{\text{1}}^{\text{rad}}$. A representative calculation of the Judd-Ofelt and related parameters, along with their related uncertainties, is given in the supporting information, S2. Across the present series, we observe the empirical ordering ($\Omega_{4} < \Omega_{2} < \Omega_{6}$), whereas some previous studies have reported ($\Omega_{4} < \Omega_{6} < \Omega_{2}$) \cite{Ebert_2025}. In particular, we see a deviation of our $\Omega_{6}$ parameters from those previously measured. For example, the Judd-Ofelt parameters were previously determined for 7-Eu in a number of solvents in addition to the solid-state \cite{Ebert_2025}. While our determined values for $\Omega_{2}$ and $\Omega_{4}$ agree with those reported for the solid-state in \cite{Ebert_2025}, the $\Omega_{6}$ parameter deviates significantly. We note that the $\Omega_{2}$ values are comparatively robust because they are determined from the intense $^{5}\text{D}_{0} \longrightarrow \ ^{7}\text{F}_{2}$ emission, while $\Omega_{6}$ is more sensitive to uncertainties in the weak long-wavelength $^{5}\text{D}_{0} \longrightarrow \ ^{7}\text{F}_{6}$ emission, detector calibration, integration limits, and the assumed refractive index. The $\Omega_{2}$ parameter is related to the local point-group symmetry around the europium ion with a higher value pertaining to a stronger $^{5}\text{D}_{0} \longrightarrow \ ^{7}\text{F}_{2}$ transition relative to the ligand-field insensitive $^{5}\text{D}_{0} \longrightarrow \ ^{7}\text{F}_{1}$ reference transition. 5-Eu and 6-Eu stand out as having particularly strong $^{5}\text{D}_{0} \longrightarrow \ ^{7}\text{F}_{2}$ transitions and therefore large $\Omega_{2}$ parameters.

Although the radiative and observed lifetimes vary significantly between the complexes, all quantum efficiencies were found to be between 50\% and 75\%, with the exception of 12-Eu, which was determined to be 23.8\%. Therefore, a significant portion of the excitation energy is lost through non-radiative pathways, namely due to the high energy C-H vibrational modes as mentioned previously. The three molecular complexes that stand out are 7-Eu, 8-Eu, and 11-Eu, with being the only three measured with quantum efficiencies above 70\%. In the context of quantum computing and communication, complexes with the lowest radiative lifetimes, while maintaining high quantum efficiencies, are favoured. 5-Eu has one of the lowest radiative lifetimes presented here of approximately 1 ms, and coupled with the optical coherence properties presented below, this complex is very promising for quantum information applications.

\begin{table*}[h!]
\small
  \caption{\ Experimentally determined Judd-Ofelt parameters, radiative ($T_{\text{1}}^{\text{rad}}$) and observed ($T_{\text{1}}^{\text{opt}}$) lifetimes and intrinsic quantum efficiencies.}
  \label{tbl3}
  \begin{tabular*}{\textwidth}{@{\extracolsep{\fill}}l|cccccc}
    \hline
    Complex & $\Omega_{2}$ [$10^{-20}$ cm$^{-2}$] & $\Omega_{4}$ [$10^{-20}$ cm$^{-2}$] & $\Omega_{6}$ [$10^{-20}$ cm$^{-2}$]& $T_{\text{1}}^{\text{rad}}$ [ms] & $T_{\text{1}}^{\text{opt}}$ [ms] & $\eta$ [\%]\\
    \hline
    1-Eu & 11.7(0.9) & 8.8(0.7) & 44.3(3.3) & 1.24(0.05) & 0.861(0.006) & 69.4(2.6)\\
    2-Eu & 15.3(1.1) & 8.5(0.6) & 33.7(2.5) & 1.04(0.04) & 0.676(0.003) & 65.0(2.6)\\
    3-Eu & 15.5(1.2) & 2.6(0.2) & 10.1(0.8) & 1.23(0.06) & 0.841(0.003) & 68.4(3.3)\\
    4-Eu & 6.4(0.5) & 5.2(0.4) & 12.8(1.0) & 2.12(0.08) & 1.317(0.015) & 62.1(2.9)\\
    5-Eu & 26.1(2.0) & 4.0(0.3) & 34.5(2.6) & 0.75(0.04) & 0.430(0.001) & 57.3(2.7)\\
    6-Eu & 26.6(2.0) & 6.2(0.5) & 37.5(2.8) & 0.71(0.03) & 0.367(0.001) & 51.7(2.4)\\
    7-Eu & 18.4(1.4) & 7.1(0.5) & 39.3(3.0) & 0.94(0.04) & 0.715(0.002) & 76.3(3.3)\\
    8-Eu & 17.3(1.3) & 7.2(0.5) & 46.1(3.5) & 0.98(0.04) & 0.698(0.001) & 71.1(3.0)\\
    9-Eu & 19.2(1.4) & 5.7(0.4) & 39.8(3.0) & 0.93(0.04) & 0.522(0.002) & 56.1(2.5)\\
    10-Eu & 8.7(0.7) & 4.6(0.3) & 29.1(2.2) & 1.73(0.07) & 1.112(0.008) & 64.3(2.7)\\
    11-Eu & 12.0(0.9) & 12.1(0.9) & 40.3(3.0) & 1.12(0.04) & 0.828(0.001) & 73.9(2.7)\\
    12-Eu & 16.7(1.3) & 2.2(0.2) & 55.3(4.2) & 1.08(0.05) & 0.257(0.002) & 23.8(1.1)\\
    \hline
  \end{tabular*}
\end{table*}

Presented in Table \ref{tbl4} are the branching ratios of the $^{5}\text{D}_{0} \longrightarrow \ ^{7}\text{F}_{J}$ ($J = 0, \cdots, 6$) transitions determined using the Judd-Ofelt parameters (J-O), where non-zero, and directly from the ratios of the integrated intensity of the multiplet of interest, over the total intensity (Dir.), as:

\begin{equation}
  \beta_{J} = \frac{A(^{7}\text{F}_{J})}{\sum_{i=0}^{6} A(^{7}\text{F}_{J})}.
  \label{Eq2}
\end{equation}

Here, $A(^{7}\text{F}_{J})$ is the integrated counts of multiplet $^{7}\text{F}_{J}$. The branching ratio of 11-Eu could not be accurately determined due to the presence of multiple sites \cite{Kuppusamy_2021}, however, the average branching ratios can still be estimated from the reported spectrum \cite{Kuppusamy_2021}. Five complexes were found to have branching ratios $\beta_0$ of over 1\%; 1-Eu, 2-Eu, 5-Eu, 8-Eu, and 11-Eu.

\begin{table*}[h!]
\small
  \caption{\ Branching ratios of the $^{5}\text{D}_{0} \longrightarrow \ ^{7}\text{F}_{J}$ (J = 0, $\cdots$, 6) transitions, calculated from the determined Judd-Ofelt parameters (J-O), where non-zero, and directly from the obtained spectra (Dir.).}
  \label{tbl4}
  \begin{tabular*}{\textwidth}{@{\extracolsep{\fill}}l|ccccccccccccccccc}
    \hline
     & $\beta_{0}$ [\%] && \multicolumn{2}{c}{$\beta_{1}$ [\%]} && \multicolumn{2}{c}{$\beta_{2}$ [\%]} && $\beta_{3}$ [\%] && \multicolumn{2}{c}{$\beta_{4}$ [\%]} && $\beta_{5}$ [\%] && \multicolumn{2}{c}{$\beta_{6}$ [\%]}\\\cline{2-2}\cline{4-5}\cline{7-8}\cline{10-10}\cline{12-13}\cline{15-15}\cline{17-18}
    Complex & Dir. && J-O & Dir. && J-O & Dir. && Dir. && J-O & Dir. && Dir. && J-O & Dir.\\
    \hline
    1-Eu & 1.28 && 8.44 & 7.78 && 61.82 & 56.92 && 4.62 && 23.64 & 21.76 && 1.62 && 6.10 & 5.61 \\
    2-Eu & 1.30 && 7.18 & 6.62 && 69.31 & 63.73 && 4.79 && 19.49 & 17.92 && 1.39 && 4.02 & 3.69 \\
    3-Eu &  --- && 8.47 & 7.86 && 83.37 & 77.20 && 5.88 && 6.75 & 6.25 && 0.68 && 1.41 & 1.31 \\
    4-Eu & 0.76 && 14.65 & 13.83 && 58.90 & 55.57 && 3.67 && 23.37 & 22.04 && 0.87 && 3.08 & 2.91 \\
    5-Eu & 1.49 && 5.19 & 4.75 && 85.52 & 78.08 && 5.76 && 6.29 & 5.75 && 0.90 && 2.99 & 2.73 \\
    6-Eu & 0.62 && 4.92 & 4.58 && 82.63 & 76.63 && 4.67 && 9.35 & 8.68 && 1.15 && 3.10 & 2.87 \\
    7-Eu & 0.37 && 6.44 & 6.14 && 75.41 & 71.66 && 3.27 && 13.89 & 13.21 && 0.97 && 4.25 & 4.04 \\
    8-Eu & 1.36 && 6.69 & 6.11 && 73.23 & 66.67 && 5.68 && 14.85 & 13.54 && 1.33 && 5.22 & 4.75 \\
    9-Eu & 0.28 && 6.40 & 5.90 && 78.19 & 72.01 && 5.67 && 11.14 & 10.25 && 1.41 && 4.27 & 3.93 \\
    10-Eu & 0.33 && 11.88 & 11.08 && 64.78 & 60.37 && 4.38 && 17.69 & 16.49 && 1.63 && 5.65 & 5.26 \\
    11-Eu & 1.11 && 7.75 & 7.22 && 58.01 & 53.72 && 3.04 && 28.52 & 26.42 && 2.92 && 5.73 & 5.38 \\
    12-Eu & 0.60 && 7.39 & 6.42 && 80.83 & 70.02 && 10.01 && 4.90 & 4.25 && 1.31 && 6.88 & 5.96 \\
    \hline
  \end{tabular*}
\end{table*}

\subsection{Optical coherence properties of 5-Eu}

\begin{figure*}[h!]
 \centering
 \includegraphics[width=1\textwidth]{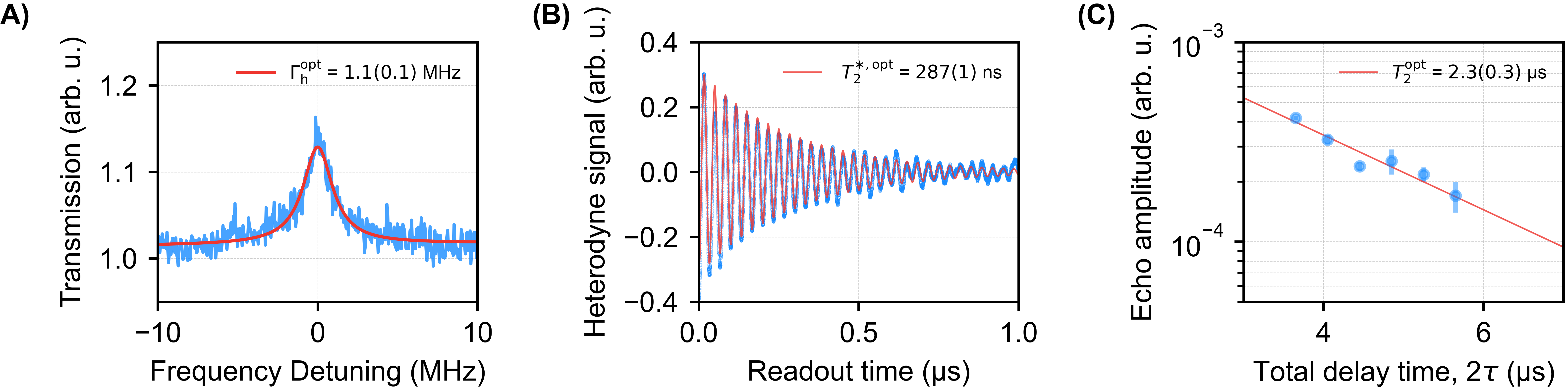}
 \caption{Spectral hole burning (A), free-induction decay (B) and two-pulse photon-echo (C) measurements of 5-Eu, measured at a temperature of 4.2 K.}
 \label{fgr7}
\end{figure*}

For Eu$^{3+}$-based molecular complexes such as those studied here to be useful in quantum computing and communication applications, they must combine a sufficiently strong $^{5}\text{D}_{0} \longrightarrow \ ^{7}\text{F}_{0}$ transition with long optical- and spin-coherence times. Although optical-coherence measurements have previously been reported for several of the molecular complexes investigated here \cite{serrano_ultra_2022, Vasilenko2026-vu, Schlittenhardt_2024, Kuppusamy_2021, Kuppusamy_2023, vasilenko2026_BB}, no such measurements have been reported for 1-Eu through 6-Eu. We therefore investigated their optical-coherence properties. Of these six complexes, only 5-Eu exhibited a measurable optical coherence signal at 4.2 K. Figure \ref{fgr7} (A) shows a spectral-hole burning (SHB) measurement performed at an excitation power of $\sim$1 mW. The fitted homogenous linewidth was $1.1(0.1)$ MHz, corresponding to an optical-coherence time of $T_{\text{2}}^{*, \text{opt}} = 1/(\pi\Gamma_{\text{h}})= 289(26)$ ns. This value agrees with the $T_{\text{2}}^{*, \text{opt}}=287(1)$ ns obtained from a free-induction decay (FID) measurement at an excitation power of 25 mW (Figure \ref{fgr7} (B)). A two-pulse photon echo measurement performed at the same excitation power (Figure \ref{fgr7} (C)) yielded a longer optical-coherence time of $T_{\text{2}}^{\text{opt}} = 2.3(0.3)$ $\upmu\mathrm{s}$, corresponding to a homogeneous linewidth of $138(18)$ kHz. These values are comparable to, or exceed, those reported for other molecular complex measured at 4.2 K. The closest comparison is 9-Eu, for which $\Gamma_{\text{h}}^{\text{opt}} =  0.31$ MHz and $T_{\text{2}}^{\text{opt}} = 2.13$ $\upmu\mathrm{s}$ were reported at 4.2 K \cite{Vasilenko2026-vu}. Measurements below 1 K are expected to yield still longer coherence times, as demonstrated for 9-Eu, where a factor of five increase was observed \cite{serrano_ultra_2022, Vasilenko2026-vu}. This is due to the significant impact of low-frequency molecular vibrational modes that get activated above $\sim 3~$K \cite{serrano_ultra_2022}.

\subsection{Discussion and suitability of REI-based molecular crystals for quantum information applications}

Table \ref{tbl5} summarises the optical- and spin-properties of the coherent $^{5}\text{D}_{0} \longrightarrow \ ^{7}\text{F}_{0}$ transition in the molecular
complexes investigated here and in previous studies. To date, optical coherence has been observed in only a subset of these complexes \cite{Kuppusamy_2021, Kuppusamy_2023, serrano_ultra_2022, Schlittenhardt_2024, Vasilenko2026-vu, vasilenko2026_BB}, while 9-Eu is, to our knowledge, the only molecular complex for which spin coherence has also been reported \cite{Vasilenko2026-vu}. Both the optical and spin coherence times are fundamentally limited by their respective population lifetimes ($T_{\text{2}} = 2T_{\text{1}}$). The values listed in Table \ref{tbl5} are, however, generally much shorter than this lifetime limit. This indicates that dephasing processes, including spectral diffusion, coupling to lattice vibrations and low-frequency molecular motions, and interactions with paramagnetic impurities, remain important in these systems. Consequently, there is considerable scope for improving the coherence times through materials design and improved measurement protocols.

\begin{table*}[h!]
\begin{threeparttable}
\small
  \caption{\ Optical and spin properties of the coherent $^{5}\text{D}_{0} \longrightarrow \ ^{7}\text{F}_{0}$ transition for all molecular complexes presented here. All measurements were performed on powder samples at 4.2 K unless otherwise specified. Fields marked with '---' have not been measured. Optical homogenous linewidths and optical-coherence times marked with a (P) were measured using a two-pulse photon echo, whereas those marked with an (S) were measured using spectral-hole burning.}
  \label{tbl5}
  \begin{tabular*}{\textwidth}{@{\extracolsep{\fill}}l|cccccccc}
    \hline
    Complex & $\beta_{0}$ [\%] & $\Gamma_{\text{inh}}^{\text{opt}}$ [GHz] & $T_{\text{1}}^{\text{opt}}$ [ms] & $\Gamma_{\text{h}}^{\text{opt}}$ [MHz] & $T_{\text{2}}^{\text{opt}}$ [$\upmu\mathrm{s}$] & $T_{\text{1}}^{\text{spin}}$ [s] & $T_{\text{2}}^{\text{spin}}$ [ms] & Ref.\\
    \hline
    1-Eu & 1.28 & 184 \& 81$^{\dag}$ & 0.86$^{\dag}$ & --- & --- & --- & --- & This study\\
    2-Eu & 1.30 & 191 \& 329$^{\dag}$ & 0.68$^{\dag}$ & --- & --- & --- & --- & This study\\
    3-Eu & --- & --- & 0.84$^{\dag}$ & --- & --- & --- & --- & This study\\
    4-Eu & 0.76 & 24.8 & 1.32$^{\dag}$ & --- & --- & --- & --- & This study\\
    5-Eu & 1.49 & 3.2 & 0.43$^{\dag}$ & 1.1 (S) & 2.3 (P) & --- & --- & This study\\
    6-Eu & 0.62 & 112 \& 196 & 0.37$^{\dag}$ & --- & --- & --- & --- & This study\\
    7-Eu & 0.37 & 12.0$^{\P}$ & 0.82$^{\text{\ss}}$ & 0.21$^{\P}$ (S) & 9.0$^{\P}$ (P) & 4.3 \& >100$^{\P}$ & --- & \cite{vasilenko2026_BB}\\
    8-Eu & 1.36 & 20.0$^{\P}$ & 0.79$^{\text{\ss}}$ & 0.24$^{\P}$ (S) & 4.0$^{\P}$ (P) & 2.7 \& 180$^{\P}$ & --- & \cite{vasilenko2026_BB}\\
    9-Eu & 0.28 & 1.9$^{\ddag}$ & 0.54* & 0.03* (P)  & 10.5* (P) & 4.4 \& 120$^{\ddag}$ & 2.0$^{\ddag}$ & \cite{serrano_ultra_2022, Vasilenko2026-vu}\\
    10-Eu & 0.33 & 16 & 1.4 & 1.55 (S) & 0.21 (S) & 0.3 \& 41 & --- & \cite{Schlittenhardt_2024}\\
    11-Eu  & 1.11 & 50* & 0.88 & 22* (S) & 0.015* (S) & 1.6 \& >30* & --- & \cite{Kuppusamy_2021} \\
    12-Eu & 0.60 & 91 & 0.35$^{\S}$ & 2.8 (S) & 0.11 (S) & 7 \& 460 & --- & \cite{Kuppusamy_2023} \\
    \hline
  \end{tabular*}

\begin{tablenotes}
\item[\dag] These measurements were performed at room temperature.
\item[\P] These measurements were performed at 100 mK.
\item[\ss] These measurements were performed at 3.0 K.
\item[\ddag] These measurements were performed on a macroscopic single crystal sample.
\item[*] These measurements were performed at 1.4 -- 1.45 K.
\item[\S] These measurements were performed at 2.3 K.

\end{tablenotes}
  
\end{threeparttable}
\end{table*}

The optical coherence is highly sensitive to temperature, and for the example of 9-Eu, a five-fold decrease in optical-coherence between 1.4 K and 4.2 K was observed \cite{serrano_ultra_2022, Vasilenko2026-vu}. Below approximately 3.5 K, dephasing was attributed primarily to coupling to two-level systems, resulting in a linear temperature dependence. At higher temperatures, quasi-localised low-frequency modes were found to dominate, leading to an approximately exponential reduction in coherence with increasing temperature \cite{serrano_ultra_2022}. Other molecular complexes are expected to exhibit similar behaviour, although the transition temperature will depend on molecular structure and crystal environment, and may vary in the range of $\sim$2 - 5 K. The measurement temperature and thermalisation conditions should therefore be considered when comparing the values in Table \ref{tbl5}. The measurements on 5-Eu and 9-Eu were performed in bath cryostats at 4.2 and 1.4 K, respectively, whereas 7-Eu and 8-Eu were measured in a dilution refrigerator at a base temperature of 100 mK. For the latter measurements, thermalisation of the molecular samples to the reported temperature could not be independently verified.

In addition, Eu$^{3+}$ concentration can strongly influence optical coherence. For example, an almost seven-fold increase in optical coherence was observed when the Eu$^{3+}$ concentration was decreased from 100\% to 25\% in ref.~\cite{serrano_ultra_2022}. Furthermore, photon-echo studies can more accurately determine the optical-coherence, as effects of power broadening, spectral diffusion and population relaxation ($T_{\text{1}}^{\text{opt}}$) can be minimised. Comparisons between complexes should therefore be regarded as qualitative unless temperature, concentration, excitation power, and measurement protocol are controlled.

\begin{figure}[h!]
 \centering
 \includegraphics[width=0.48\textwidth]{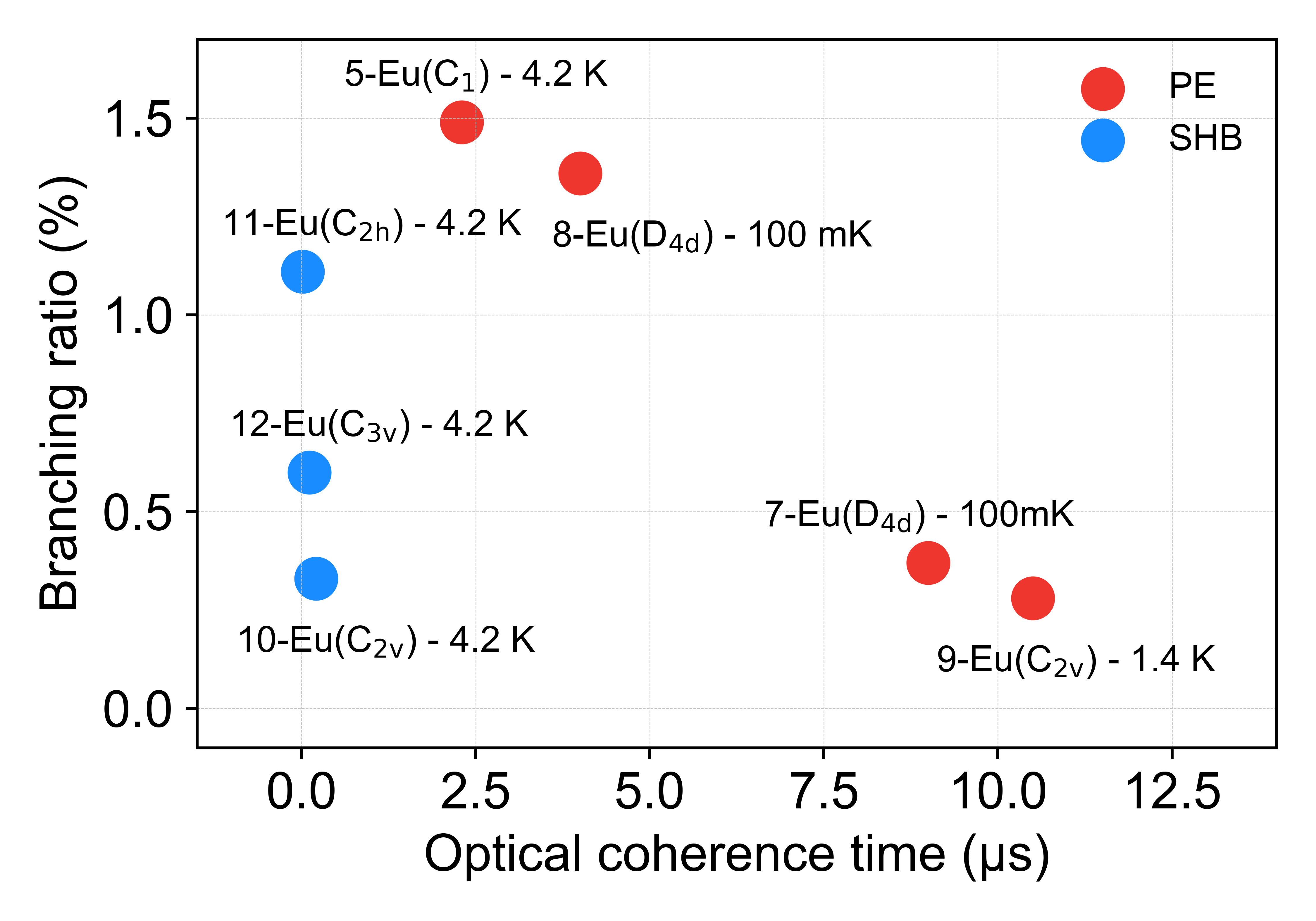}
 \caption{Comparison of optical coherence, point-group symmetry and branching ratios of all molecular complexes with measurable optical coherence investigated in this study. Data points in red were measured using two-pulse photon-echo methods, whereas those in blue were measured using spectral-hole burning.}
 \label{fgr8}
\end{figure}

Despite these limitations, some trends can be established. Four complexes have shown optical-coherence times greater than 1 $\upmu\mathrm{s}$ (5-Eu, 7-Eu, 8-Eu and 9-Eu). Of these, 5-Eu and 8-Eu have $^{5}\text{D}_{0} \longrightarrow \ ^{7}\text{F}_{0}$ branching ratios above 1\%, an order of magnitude larger than 7-Eu and 9-Eu. Figure \ref{fgr8} compares the branching ratio and the measured optical coherence time. Data points in red were measured using photon-echo methods, whereas those in blue were measured using spectral-hole burning. Visible are multiple trends. Firstly, as mentioned above, photon-echo studies results in higher measured coherence times. The data separate approximately into two groups, with branching ratios approximately below 0.5\% and above 1\%. The former group contains complexes assigned to C$_{\text{2v}}$ or higher symmetry, whereas the latter predominantly comprises complexes with lower point-group symmetries of either C$_{2}$ or C$_{1}$. This is consistent with the increased $J$-mixing expected in lower-symmetry ligand fields, which can lend intensity to the nominally
forbidden $^{5}\mathrm{D}_{0}\rightarrow{}^{7}\mathrm{F}_{0}$ transition. Also visible, for all complexes where the coherence times were measured via photon-echo studies is that complexes with higher point-group symmetries (7-Eu and 9-Eu) have higher coherence times than those with lower symmetries (5-Eu and 8-Eu). As mentioned previously, the optical coherence time of 5-Eu was measured at 4.2 K; therefore, this value is expected to improve at lower temperature. This could make 5-Eu potentially have the best combination of optical coherence time and branching ratio investigated in this study.

For quantum-information applications, the intrinsic weakness of the $^{5}\mathrm{D}_{0}\rightarrow{}^{7}\mathrm{F}_{0}$ transition presents a challenge. Increasing its oscillator strength through chemical modification is difficult because the transition is doubly forbidden and its intensity
depends on weak intensity-borrowing mechanisms. Optical resonators provide a complementary route by selectively enhancing the emission rate through the Purcell effect.
In earlier work, the complex \ce{[Eu(TTA)3(DPEPO)]}, where TTA denotes Thenoyltrifluoroacetone and DPEPO denotes \ce{bis(2-(diphenylphosphino)phenyl)} ether oxide, has been integrated into both a Fabry-P\'erot cavity \cite{Ruggero_2022} and a nanobeam cavity \cite{Ruggero_2023}. In the nanobeam cavity, a 140-fold increase in the emission rate of the $^{5}\text{D}_{0} \longrightarrow \ ^{7}\text{F}_{2}$ transition was observed. Similarly, a Purcell factor of 410 was measured for the coherent $^{5}\text{D}_{0} \longrightarrow \ ^{7}\text{F}_{0}$ transition when 8-Eu was integrated into a Fabry-P\'erot cavity \cite{vasilenko2026_BB}. This resulted in a reported increase of the branching ratio from 1.30\% to 72\%. However, with the measured Purcell enhancement and the comparably short coherence time of the complex, the system is still a factor of $\sim$70 away from achieving lifetime-limited optical coherence ($T_{\text{2}} = 2T_{\text{1}}$), an important requirement for spin-photon interfaces. Further improvements of materials and cavities promise to achieve this regime.

\section{Conclusion}
We have presented a systematic optical study of twelve Eu$^{3+}$-based molecular complexes, including their emission branching ratios, excited-state lifetimes, quantum efficiencies, Judd--Ofelt parameters, and optical- and
spin-coherence properties. The results demonstrate that the suitability of these materials for quantum-information applications depends on several interrelated, but distinct, properties.

The branching ratio of the nominally forbidden $^{5}\mathrm{D}_{0}\rightarrow{}^{7}\mathrm{F}_{0}$ transition shows a qualitative dependence on effective site symmetry. Complexes with lower effective symmetries, particularly $C_{2}$ and $C_{1}$, generally exhibit larger branching ratios than complexes with $C_{2v}$ or higher symmetry. This is consistent with increased mixing between different $J$ manifolds in lower-symmetry ligand fields. The point-group assignment alone is not sufficient, however, since distortions from ideal geometries, donor-atom arrangements, Eu--ligand distances, and the magnitude and polarity of the local ligand field also influence the transition intensity.

The Judd--Ofelt parameters provide complementary information about the local ligand-field environment. In particular, $\Omega_{2}$ reflects the hypersensitivity of the $^{5}\mathrm{D}_{0}\rightarrow{}^{7}\mathrm{F}_{2}$ transition and is related to contributions from local asymmetry, polarisability, and covalency.

The coherence measurements show that a large $^{5}\mathrm{D}_{0}\rightarrow{}^{7}\mathrm{F}_{0}$ branching ratio does not necessarily correlate with long optical coherence. Of the newly investigated complexes 1-Eu -- 6-Eu, only 5-Eu exhibited a measurable coherence signal at 4.2 K, with a photon-echo coherence time of $2.3(0.3)~\upmu\mathrm{s}$. Optical coherence is also strongly affected by temperature, Eu$^{3+}$ concentration, excitation power, measurement protocol, crystal packing, and low-frequency molecular dynamics.

These results suggest that molecular design should combine a suitably asymmetric ligand field with a rigid and structurally homogeneous environment. High-energy vibrational modes should be minimised to reduce non-radiative multiphonon relaxation, while crystal packing and Eu$^{3+}$ -- Eu$^{3+}$ interactions must be controlled to limit spectral diffusion and dephasing. Optical cavities provide a complementary route to enhancing the weak $^{5}\mathrm{D}_{0}\rightarrow{}^{7}\mathrm{F}_{0}$ transition without relying solely on chemical modification. Together with first demonstrations of ion-ion interactions \cite{serrano_ultra_2022,vasilenko2026_BB} and compatibility with atomic-frequency-comb protocols \cite{serrano_ultra_2022}, these results support Eu$^{3+}$ molecular crystals as promising platforms for integrated quantum memories and spin--photon interfaces. Future temperature-dependent coherence measurements, single-crystal structural characterisation, and quantitative crystal-field calculations will be important for establishing predictive design rules.

\section*{Author contributions}

\textbf{N.L. Jobbitt}: Investigation, Formal analysis, Supervision, Writing - original draft, Writing - review \& editing. \textbf{S. M. Weis}: Investigation, Formal analysis, Writing - original draft, Writing - review \& editing. \textbf{T. Schweizer}: Synthesis, Writing - original draft, Writing - review \& editing. \textbf{V. Unni Chorakkunnath}: Investigation, Formal analysis, Writing - original draft, Writing - review \& editing. \textbf{E. Vasilenko}: Investigation, Formal analysis, Writing - original draft, Writing - review \& editing. \textbf{T. Haas}: Synthesis, Writing - original draft, Writing - review \& editing.  \textbf{B. Brachnakova}: Synthesis, Writing - original draft, Writing - review \& editing. \textbf{S.K. Kuppusamy}: Synthesis, Investigation, Formal analysis, Supervision, Writing - original draft, Writing - review \& editing. \textbf{H. Schubert}: X-ray diffraction, Investigation, Formal analysis, Writing - original draft, Writing - review \& editing.  \textbf{M. Ruben}: Conceptualisation, Supervision, Formal analysis, Writing - review \& editing. \textbf{M. Seitz}: Conceptualisation, Supervision, Formal analysis, Writing - review \& editing. \textbf{D. Hunger}: Conceptualisation, Supervision, Project administration, Writing - original draft, Writing - review \& editing.

\section*{Conflicts of interest}
There are no conflicts to declare.

\section*{Data availability}

The data supporting this article have been included as part of the Supplementary Information.


\section*{Acknowledgements}

The authors acknowledge funding from the Deutsche Forschungsgemeinschaft (DFG) through the Collaborative Research Centre “4f for Future” (CRC 1573 project number 471424360, project C2), the Max-Planck-School of Photonics, and the Karlsruhe School of Optics and
Photonics (KSOP). M. R. acknowledges funding from the government programme managed by the French National Research Agency under grants ANR-20-CE09-0022-01 (UltraNanoSpec), ANR-23-CE47-0011 (MoleQuBe) and France 2030-ANR-MOlQif (PEPR Technologies Quantiques).



\balance

\renewcommand\refname{References}

\bibliography{bibliography} 
\bibliographystyle{rsc} 

\end{document}